\documentclass[aps,pra,superscriptaddress,floatfix,showpacs,10pt,two
column]{revtex4}%
\usepackage[dvips]{graphicx}
\usepackage{amssymb}
\usepackage{amsmath}
\usepackage{color}
\usepackage{amsfonts}%
\definecolor{ueblue}{rgb}{0,0,0.2}
\definecolor{hh}{rgb}{1,0.4,0.3}
\begin{document}
\title{Scheme for the implementation of optimal cloning of arbitrary
single particle atomic state into two photonic states}
\author{Wei Song}
\email{wsong1@mail.ustc.edu.cn}
\affiliation{Hefei National Laboratory for Physical Sciences at
Microscale and Department of Modern Physics, University of Science and Technology of China, Hefei,
Anhui 230026, China}
\author{Tao Qin}

\affiliation{Hefei National Laboratory for Physical Sciences at Microscale and Department of Modern
Physics, University of Science and Technology of China, Hefei, Anhui 230026, China}


\begin{abstract}
We present a feasible scheme to implement the $1 \to 2$ optimal cloning of arbitrary single
particle atomic state into two photonic states, which is important for applications in long
distance quantum communication.  Our scheme also realizes the tele-NOT gate of one atom to the
distant atom trapped in another cavity. The scheme is based on the adiabatic passage and the
polarization measurement. It is robust against a number of practical noises such as the violation
of the Lamb-Dicke condition, spontaneous emission and detection inefficiency.
\end{abstract}
\maketitle

One of the most fundamental differences between classical and quantum information is that while
classical information can be copied perfectly, quantum information cannot. In particular, it
follows from the no-cloning theorem\cite{Wootters:1982} that one cannot create a perfect duplicate
of an arbitrary state. Although perfect cloning is not allowed, it is, nevertheless, possible to
construct approximate\cite{Buzek:1996} or probabilistic cloning\cite{Duan:1998}machine. The
approximate cloning machine transforms the arbitrary input states into imperfect copies with
probability one. While in probabilistic cloning one always obtain a perfect copy with some
probability less than one. The simplest cloning machine is the duplication of a qubit, as was
considered in \cite{Buzek:1996}. Many generalizations and variants have followed, such as $N \to M$
optimal universal cloning machine for qubits\cite{Gisin:1997}, or d-level
systems\cite{Werner:1998}, state-dependent cloning machine\cite{Bruss:1998}, phase-covariant
cloning machine for equatorial qubits\cite{Fan:2002}, cloning machine for
continuous-variable\cite{Cerf:2000} systems etc. $1 \to 2$ and $1 \to 3$ universal cloning machines
are also extended to the asymmetric case\cite{Cerf1:2000}. The study of quantum cloning has
increased our understanding of the properties of quantum information. It has been shown that
quantum cloning has close connection to the assessment of security\cite{Fuchs:1997} in quantum
cryptography. Applications of quantum cloning can also be found in many quantum information
\cite{Ricci:2005} and quantum computation tasks\cite{Galvao:2000}. All these motivations have led
to the rapid development in the theory studies of quantum cloning\cite{Scarani:2005}.

On the other hand, it is important to find a specific physical means to carry out a given cloning
process. Several schemes for realization of different quantum cloning processes have been suggested
with quantum optics\cite{Simon:2000}, linear optics\cite{Fiurasek:2001}, cavity
QED\cite{Milman:2003} and spin networks\cite{Chiara:2004}. Recently there have been greatly
progresses in experiment for demonstrating the various kinds of cloning
machines\cite{Huang:2001,Ricci:2004}. In Ref.\cite{Fiurasek:2004} a scheme for continuous variable
cloning of light into an atomic ensembles has been proposed. However, the inverse process is also
important because the photonic state is more suitable for long-distance communication. In this
paper, we will propose a scheme to implement the $\mbox{1} \to \mbox{2}$ optimal cloning of
arbitrary single particle atomic state into two photonic states. Our scheme combines the cavity QED
technology and simple linear optical elements. The cavity QED system \cite{Raimond:2001} is a
promising candidate for quantum information processing. In cavity QED, the atoms act as the
stationary qubits and they are coupled via interaction with the cavity photons. Our scheme is a
combination of the two advantages: atom acts as stationary qubit used only for memory, while
photons play the role of flying qubits. Our scheme also realizes the tele-NOT gate of one atom to
the distant atom trapped in another cavity.

\begin{figure}[ptb]
\includegraphics[scale=0.55,angle=0]{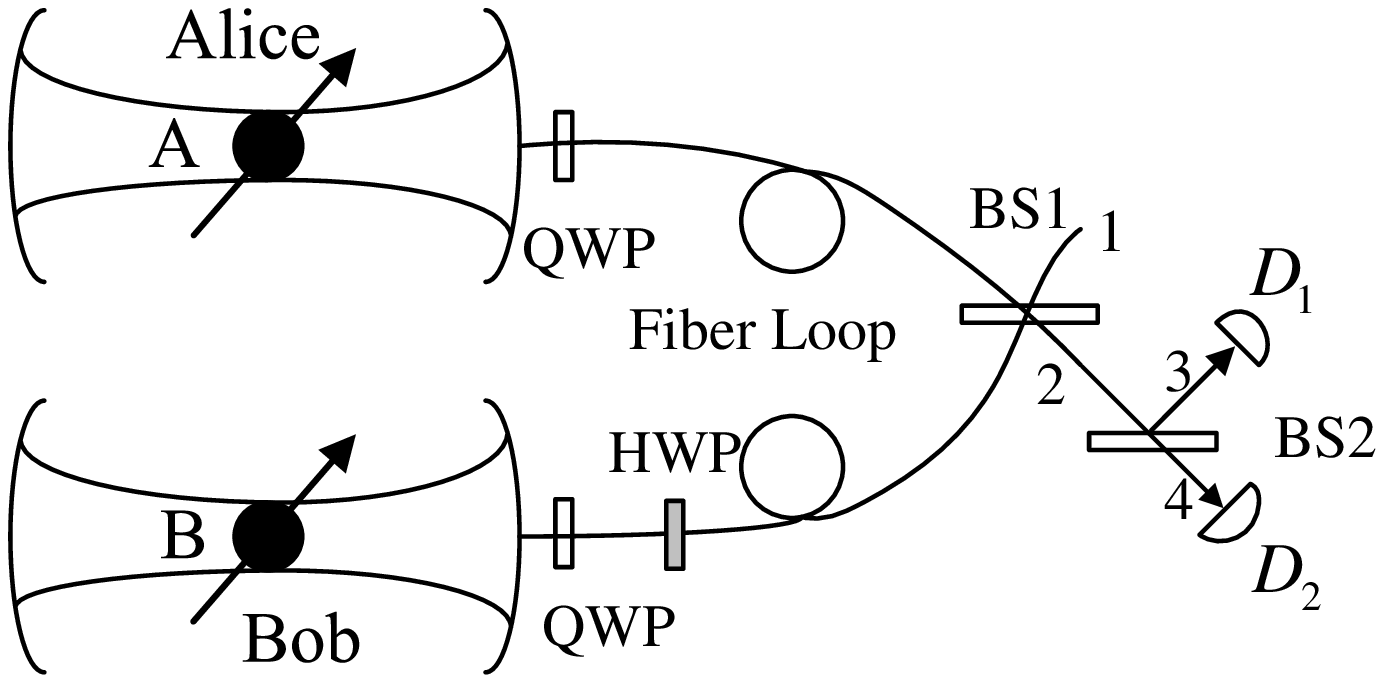}\caption{Schematic setup to implement the  $1 \to 2$  optimal symmetric cloning of arbitrary single particle atomic state into two photonic states. Here BS denotes beam splitter, QWP denotes quarter-wave plate and HWP is a $0^ \circ $ half wave plate.}%
\label{fig1}%
\end{figure}

Before presenting our scheme, let us review the $\mbox{1} \to \mbox{2}$ optimal cloning process
firstly. Suppose the unknown single-qubit state to be cloned is in the form $\left| \psi
\right\rangle _1 = a\left| 0 \right\rangle _1 + b\left| 1 \right\rangle _1 $, where $\left| a
\right|^2 + \left| b \right|^2 = 1$. The ancillary qubits are in a singlet state $\left| \psi
\right\rangle _{23}^ - = 2^{ - \frac{1}{2}}\left( {\left| {01} \right\rangle _{23} - \left| {10}
\right\rangle _{23} } \right)$. The $1 \to 2$ optimal cloning of the state $\left| \psi
\right\rangle _1 $ corresponds to the projection operation onto the subspace of the total state
$\Pi = \left| \psi \right\rangle _1 \otimes \left| \psi \right\rangle _{23}^ - $ , and the
projector operator is given by\cite{Werner:1998}

\begin{equation}
\label{eq1} P_{123} = \left( {{\rm I}_{12} - \left| {\psi ^ - } \right\rangle _{12} \left\langle
{\psi ^ - } \right|_{12} } \right) \otimes {\rm I}_3 .
\end{equation}

The procedure above will generate the state $\tilde {\Pi } = \sqrt {\frac{2}{3}} \left( {\left|
{\zeta _1 } \right\rangle _{12} \left| 1 \right\rangle _3 - \left| {\zeta _0 } \right\rangle _{12}
\left| 0 \right\rangle _3 } \right)$, where $\left| {\zeta _1 } \right\rangle _{12} = a \left| {00}
\right\rangle _{12} + \frac{1}{2}b \left( {\left| {01} \right\rangle _{12} + \left| {10}
\right\rangle _{12}}  \right)$, and $\left| {\zeta _0 } \right\rangle _{12} = b \left| {11}
\right\rangle _{12} + \frac{1}{2}a \left( {\left| {01} \right\rangle _{12} + \left| {10}
\right\rangle _{12} } \right)$. Here, the two cloning states emerge in 1 and 2 qubits. By tracing
out the qubits 1 and 2, we find the qubit 3 realize optimal universal-NOT(UNOT) gate.

In the following we will give our scheme in detail. The system we are considering consists of two
optical cavities with atoms A is trapped in Alice's cavity and the atom B in Bob's cavity as shown
in Fig.1. The level structures of the atom A and B are depicted in Fig.2. Alice and Bob exploit two
$F = 1$ hyperfine levels, while Bob exploits one additional hyperfine level. Such atomic level
structures can be achieved in $^{87}Rb$. The states $\left( {5{ }^2S_{\frac{1}{2}} ,F = 1} \right)$
and $\left( {5{ }^2P_{\frac{3}{2}} ,F = 1} \right)$ correspond to the $F = 1$ ground and excited
hyperfine levels,respectively. The states $\left( {5{ }^2S_{\frac{1}{2}} ,F = 2,m = 0} \right)$
correspond to $\left| {{g}'_0 } \right\rangle _B $. In our scheme, the atom A to be cloned is
encoded in the superposition state of $\left| {g_L } \right\rangle $ and $\left| {g_R }
\right\rangle $. With cavity A prepared in the vacuum state, the initial state of the whole system
of Alice is

\begin{equation}
\label{eq2} \left| {\psi \left( 0 \right)} \right\rangle _A = \left( {a\left| {g_L } \right\rangle
_A + b\left| {g_R } \right\rangle _A } \right)\left| {0,0} \right\rangle _A ,
\end{equation}

\noindent where we have used the notation $\left| {n_{L,} n_R } \right\rangle _i $ , $n_{L,R} $
represents the number of left- or right-circularly polarized photons and $i = A,B$ represents Alice
or Bob, respectively. The transitions $\left| {g_m } \right\rangle _A \to \left| {e_m }
\right\rangle _A \left( {m = L,R} \right)$ are driven adiabatically through the laser pulse
collinear with the cavity axis, and atom A will be transferred with probability $P_1 \simeq 1$ to
the state $\left| {g_0 } \right\rangle _A $ by emitted a photon from the transitions $\left| {e_L }
\right\rangle _A \to \left| {g_0 } \right\rangle _A $ and $\left| {e_R } \right\rangle _A \to
\left| {g_0 } \right\rangle _A $. The corresponding Rabi frequency of the laser pulse is $\Omega _A
\left( t \right)$; the transitions $\left| {e_L } \right\rangle _A \to \left| {g_0 } \right\rangle
_A $ and $\left| {e_R } \right\rangle _A \to \left| {g_0 } \right\rangle _A $ are coupled to the
left-circularly and right-circularly polarized mode of the cavity with the coupling rate $g_A $. In
the rotating frame, the Hamiltonian of Alice's system is given by \cite{Cho:2004}

\begin{equation}
\label{eq3}
\begin{array}{l}
 H_A = - \left( {\Delta + i\frac{\gamma _A }{2}} \right)\left( {\left| {e_L
} \right\rangle \left\langle {e_L } \right| + \left| {e_R } \right\rangle \left\langle {e_R }
\right|} \right)_A + \left[ {\Omega _A \left( t \right)}
\right.\left| {e_L } \right\rangle \left\langle {g_L } \right| \\
 \left. {\left. {\left| {e_R } \right\rangle \left\langle {g_R } \right|}
\right)_A + g_A \left( {a_L^A \left| {e_L } \right\rangle \left\langle {g_0 } \right| + a_R^A
\left| {e_R } \right\rangle \left\langle {g_0 } \right|}
\right)_A + H.c.} \right], \\
 \end{array}
\end{equation}

\noindent where $\gamma _A $ and $a_{L,R}^A $ denote the atomic spontaneous emission rate and the
annihilation operator for the corresponding polarized mode of the cavity, respectively. The
Hamiltonian has two orthogonal dark state$\left| {D_1 \left( t \right)} \right\rangle _A = \cos
\theta _A \left( t \right)\left| {g_L } \right\rangle _A \left| {0,0} \right\rangle _A - \sin
\theta _A \left( t \right)\left| {g_0 } \right\rangle _A \left| {1,0} \right\rangle _A $ and
$\left| {D_2 \left( t \right)} \right\rangle _A = \cos \theta _A \left( t \right)\left| {g_R }
\right\rangle _A \left| {0,0} \right\rangle _A - \sin \theta _A \left( t \right)\left| {g_0 }
\right\rangle _A \left| {0,1} \right\rangle _A $ with $\cos \theta _A = \frac{g_A }{\sqrt {\left|
{g_A } \right|^2 + \left| {\Omega _A } \right|^2} }$ and $\sin \theta _A = \frac{\Omega _A(t)
}{\sqrt {\left| {g_A } \right|^2 + \left| {\Omega _A } \right|^2} }$. Under the adiabatic
approximation, the initial state (\ref{eq2}) evolve into the following state:

\begin{equation}
\label{eq4}
\begin{array}{l}
 \left| {\Psi \left( t \right)} \right\rangle _A = \cos \theta _A \left( t
\right)\left( {a\left| {g_L } \right\rangle _A + b\left| {g_R }
\right\rangle _A } \right)\left| {0,0} \right\rangle _A \\
 - \sin \theta _A \left( t \right)\left| {g_0 } \right\rangle _A \left(
{a\left| {1,0} \right\rangle _A + b\left| {0,1} \right\rangle _A } \right).
\\
 \end{array}
\end{equation}

\begin{figure}[ptb]
\includegraphics[scale=1,angle=0]{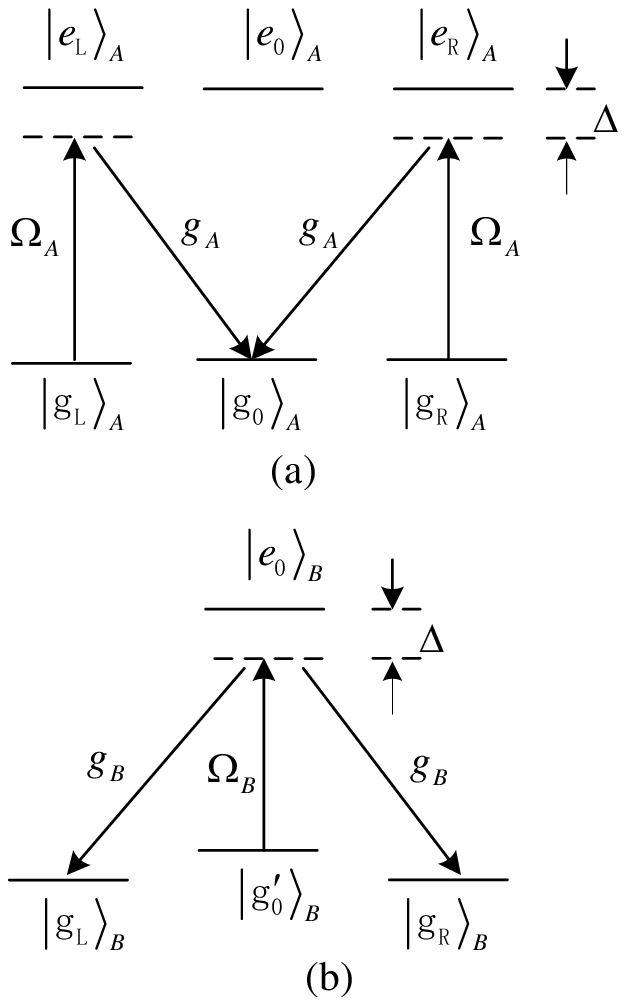}\caption{The level configuration and transitions of the atom for Alice(a) and Bob (b). }%
\label{fig2}%
\end{figure}

Alice slowly increasing $\Omega _A $ to satisfy the condition $\frac{g_A }{\Omega _A } \approx 0$
which will adiabatically transform the state $ {a\left| {g_L } \right\rangle _A + b\left| {g_R }
\right\rangle _A } $ into $ {a\left| {1,0} \right\rangle _A + b\left| {0,1} \right\rangle _A } $.

The procedure for Bob is similar to Alice. Bob's system is prepared in the state $\left| {{g}'_0 }
\right\rangle _B \left| {0,0} \right\rangle _B $. The transitions $\left| {{g}'_0 } \right\rangle
_B \to \left| {e_0 } \right\rangle _B $are driven adiabatically by laser pulse collinear with the
cavity axis. $\left| {e_0 } \right\rangle _B \to \left| {g_L } \right\rangle _B $ and $\left| {e_0
} \right\rangle _B \to \left| {g_R } \right\rangle _B $ are coupled to the right-circularly and
left-circularly polarized mode of the cavity with the coupling rate $g_B $.The atom of B will be
transferred with probability $P_2 \simeq 1$ to the state $\left| {g_L } \right\rangle _B $ and
$\left| {g_R } \right\rangle _B $ by emitted a photon from the transition $\left| {e_0 }
\right\rangle _B \to \left| {g_L } \right\rangle _B $ and $\left| {e_0 } \right\rangle _B \to
\left| {g_R } \right\rangle _B $. The Hamiltonian of Bob's system in the rotating frame is given by

\begin{equation}
\label{eq5}
\begin{array}{l}
 H_B = - \left( {\Delta + i\frac{\gamma _B }{2}} \right)\left( {\left| {e_0
} \right\rangle \left\langle {e_0 } \right|} \right)_B + \left[ {\Omega _B \left( t \right)}
\right.\left( {\left| {e_0 } \right\rangle \left\langle
{{g}'_0 } \right|} \right)_B \\
 \left. { + g_B \left( {a_R^B \left| {e_0 } \right\rangle \left\langle {g_L
} \right| + a_L^B \left| {e_0 } \right\rangle \left\langle {g_R } \right|}
\right)_B + H.c.} \right], \\
 \end{array}
\end{equation}

Under the adiabatic approximation, Bob's system will evolve into the following state:

\begin{equation}
\label{eq6}
\begin{array}{l}
 \left| {\Psi \left( t \right)} \right\rangle _B = \cos \theta _B \left( t
\right)\left| {{g}'_0 } \right\rangle _B \left| {0,0} \right\rangle _B - \frac{\sin \theta _B
\left( t \right)}{\sqrt 2 }\left( {\left| {g_L }
\right\rangle } \right._B \left| {0,1} \right\rangle _B \\
+ \left| {g_R } \right\rangle _B \left. {\left| {1,0} \right\rangle _B }
\right) \\
 . \\
 \end{array}.
\end{equation}

\noindent where $\cos \theta _B = \frac{\sqrt 2 g_B }{\sqrt {2\left| {g_B } \right|^2 + \left|
{\Omega _B } \right|^2} }$ and $\sin \theta _B = \frac{\Omega _B(t) }{\sqrt {2\left| {g_B }
\right|^2 + \left| {\Omega _B } \right|^2} }$. Bob also increase $\Omega _B $ adiabatically to map
his state into a maximally entangled state $\frac{\mbox{1}}{\sqrt 2 }\left( {\left| {g_L }
\right\rangle } \right._B \left| {0,1} \right\rangle _B + \left| {g_R } \right\rangle _B \left.
{\left| {1,0} \right\rangle _B } \right)$.

Because the cavities are one-side leaky, each cavity emits a photon and interferes at the beam
splitter 1. The quarter-wave plates transform left-polarized and right-polarized photons into
horizontally and vertically polarized photons with the transformation $\left| {1,0 } \right\rangle
\to \left| H \right\rangle $ and $\left| {0,1 } \right\rangle \to \left| V \right\rangle $, where
we have ignored the vacuum modes due to their no contribution to the click of the photondetectors.
$\left| H \right\rangle $ and $\left| V \right\rangle $ denote the horizontally and vertically
polarized photons, respectively. After the photons passing the two QWP and HWP, the total state of
Alice and Bob's system will evolves into

\begin{equation}
\label{eq7}
\begin{array}{l}
 \left| \Phi \right\rangle = \frac{1}{\sqrt 2 }\left| {g_0 } \right\rangle
_A \left( {a\left| H \right\rangle _A + b\left| V \right\rangle _A } \right)( \left| {g_R }
\right\rangle _B \left. {\left| H \right\rangle _B -
\left| {g_L } \right\rangle _B \left| V \right\rangle _B } \right) \\
 . \\
 \end{array}
\end{equation}

In order to realize the $1 \to 2$ optimal cloning of the single qubit state, the central task is to
realize the projective measurement given by Eq.(\ref{eq1}). As shown in Ref.\cite{Ricci:2004},the
projective measurement in the form of Eq.(\ref{eq1}) can be realized by the superposition of two
modes on the 50:50 beam splitter. It can be identified by the simultaneous clicking of the
detectors $D_1 $ and $D_2 $. We only consider the mode 2 because the same effect is expected on
mode 1, for simplicity, we only depict the setup in mode 2 which applies to the mode 1 also. The
setup of our scheme is depicted in Fig.1. The term $\left| {HH} \right\rangle _{AB} ,\left| {HV}
\right\rangle _{AB} ,\left| {VH} \right\rangle _{AB} ,\left| {VV} \right\rangle _{AB} $ will
undergo the following transformation if there is two-fold coincidence of $D_1 $ and $D_2 $

\begin{equation}
\label{eq8}
\begin{array}{l}
 \left| {HH} \right\rangle _{AB} \to \left| {HH} \right\rangle _{34} , \\
 \left| {HV} \right\rangle _{AB} \to \frac{1}{2}\left( {\left| {HV}
\right\rangle _{34} + \left| {VH} \right\rangle _{34} } \right), \\
 \left| {VH} \right\rangle _{AB} \to \frac{1}{2}\left( {\left| {HV}
\right\rangle _{34} + \left| {VH} \right\rangle _{34} } \right), \\
 \left| {VV} \right\rangle _{AB} \to \left| {VV} \right\rangle _{34} . \\
 \end{array}
\end{equation}

The scheme above requires the twofold coincidence event as the indication that the projector
$P_{123} $ in the form of Eq.(\ref{eq1}) has been performed. Once the two detectors click
simultaneously, the total state will evolves into

\begin{equation}
\label{eq9}
\begin{array}{l}
 \sqrt {\frac{\mbox{2}}{\mbox{3}}} a\left| {HH} \right\rangle _{34} \left|
{g_R } \right\rangle_B - \sqrt {\frac{\mbox{1}}{\mbox{6}}} a\left( {\left| {HV} \right\rangle _{34}
+ \left| {VH} \right\rangle _{34} } \right)\left|
{g_L } \right\rangle_B \\
 - \sqrt {\frac{\mbox{2}}{\mbox{3}}} b\left| {VV} \right\rangle _{34} \left|
{g_L } \right\rangle_B + \sqrt {\frac{\mbox{1}}{\mbox{6}}} b\left( {\left| {HV} \right\rangle _{34}
+ \left| {VH} \right\rangle _{34} } \right)\left|
{g_R } \right\rangle_B . \\
 \end{array}
\end{equation}

The equation above shows that the optimal $1 \to 2$ symmetric cloning process has been implemented.
This scheme also realizes the tele-NOT gate of atom A to atom B trapped in another cavity. By
taking consideration the coincidence of the mode 1, the total success probability of obtaining the
two-fold coincidence event is $\frac{\mbox{1}}{\mbox{4}}$.

Next we give a brief discussion on feasibility of our scheme. In the actual situation, each photon
leaks from the cavity in the form of single-pulse due to the random nature of the emission. The two
photons interfere maximally when the two pulse shapes overlap completely at the beam splitter. In
the adiabatic limit, the single-photon pulse shape is given by \cite{Duan:2003}$f_i \left( t
\right) = \sqrt {\kappa _i } \sin \theta _i \left( t \right)\exp \left( { - \frac{\kappa _i
}{2}\int_0^t {\sin ^2\theta _i \left( \tau \right)d\tau } } \right),$ where $\kappa _i $ denotes
the cavity decay rate for Alice or Bob. As shown in Ref.\cite{Yu:2004}, the difference between the
two pulse shapes can be small enough if we choose the appropriate driving pulse. During our
discussion we have assumed the coupling rates are fixed. However, the coupling rates have a
variation in time due to the thermal motion of the atom, thus lead to the change of the output of
the pulse shape $f_i \left( t \right)$. The result of the numerical simulation\cite{Cho:2004} shows
our scheme works beyond the restriction of the Lamb-Dicke condition and the fidelity is only
affected slightly. By utilizing the adiabatic method, the atomic decay is highly suppressed in our
scheme. Another{ problem is the influence of the imperfections of the single-photon detectors,
e.g., inefficient detections and the dark counts. If photons leak out of the cavity, but are not
detected, these processes simply decrease the success probability by a factor of $\eta^{2}$
($\eta$: the detection efficiency of the single-photon detectors), but have no influence on the
fidelity of our scheme. For real single-photon detectors the dark counts can be in the few-percent
region \cite{Blinov:2004}. However, the utility of the twofold coincidence in our scheme will
greatly suppress the dark counts. As such, the effect of the dark counts only affects the scheme
slightly and can be safely neglected.

By combining the cavity QED technology and simple linear optical elements, we have proposed a
scheme for the implementation symmetric quantum cloning of arbitrary single particle atomic state
into two photonic states. The feature of our proposal is that the state to be cloned is encoded in
the atomic state which is more suitable for memory. While the clones are appearing in two photonic
states, which is important for long distance quantum communication protocols. Furthermore, our
proposal is robust against a number of practical noises such as the violation of the Lamb-Dicke
condition, spontaneous emission and detection inefficiency.

This work was supported by the National NSF of China, the Fok Ying Tung Education Foundation, and
the Chinese Academy of Sciences.

\end{document}